\documentclass[reqno,12pt]{amsart}
\usepackage{fullpage}
\usepackage{amsfonts}
\usepackage{amssymb}

\numberwithin{equation}{section}
\def\X{{\bf X}}
\def\Y{\hat{\bf X}}
\def\pr{{\rm pr}}
\def\p{\partial}

\def\E{{\mathcal E}}
\def\triv{\text{triv.}}
\def\ext{\text{ext.}}

\def\Rnum{\mathbb{R}}

\newtheorem{prop}{Proposition}
\newtheorem{thm}{Theorem}
\newtheorem{cor}{Corollary}
\newtheorem{lem}{Lemma}

\def\propref#1{Proposition~\ref{#1}}
\def\proprefs#1#2{Propositions~\ref{#1} and~\ref{#2}}

\def\thmrefs#1#2{Theorems~\ref{#1} and~\ref{#2}}
\def\lemref#1{Lemma~\ref{#1}}
\def\Ref#1{Ref.\cite{#1}}
\def\Refs#1{Refs.\cite{#1}}
\def\secref#1{Sec.~\ref{#1}}

\def\ie/{i.e.}
\def\eg/{e.g.}
\def\etc/{etc.}

\begin{document}
\allowdisplaybreaks[3]

\title{Symmetry properties of conservation laws}

\author{
Stephen C. Anco
\\\lowercase{\scshape{
Department of Mathematics and Statistics\\
Brock University\\
St. Catharines, ON L2S3A1, Canada}} \\
}

\begin{abstract}
Symmetry properties of conservation laws of partial differential equations 
are developed by using the general method of conservation law multipliers. 
As main results, 
simple conditions are given for characterizing when a conservation law 
and its associated conserved quantity are invariant 
(and, more generally, homogeneous) under the action of a symmetry. 
These results are used to show that a recent conservation law formula 
(due to Ibragimov) is equivalent to a standard formula 
for the action of an infinitesimal symmetry on a conservation law multiplier. 
\end{abstract}

\maketitle

\section{Introduction}
\label{intro}

For systems of partial differential equations (PDEs) that 
possesses a Lagrangian formulation, 
it is well-known that conservation laws arise from variational symmetries
through Noether's theorem \cite{Olv,1stbook,2ndbook}. 
Interestingly, 
variational symmetries have an equivalent characterization as multipliers
\cite{Olv,2ndbook,BluCheAnc}, 
which comes from expressing the symmetry generator in an evolutionary form
acting only on the dependent variable(s) in the given PDE system. 
This characterization has the important advantage that 
it is able to provide an alternative method to Noether's theorem 
to find all conservation laws, without using a Lagrangian. 
The method uses a linear system of determining equations 
whose solutions yield all multipliers and hence all variational symmetries. 
This determining system can be solved by the same standard procedure 
used to solve the determining equations for symmetries. 
For each multiplier, a corresponding conservation law can then be obtained 
by various direct integration methods \cite{2ndbook,BluCheAnc,Anc}, 
hence by-passing the need to use the Lagrangian 
(as well as the need to consider ``gauge terms'' in the definition of variational symmetries). 

Most significantly, 
this direct method using multipliers has a straightforward extension 
\cite{AncBlu97,AncBlu02a,AncBlu02b,review}
to any PDE system of normal type
regardless of whether a Lagrangian exists for the system. 
In general the determining system for multipliers 
consists of the adjoint of the symmetry determining equations 
plus additional equations analogous to Helmholtz conditions. 
The standard procedure for solving the symmetry determining equations 
\cite{Olv,1stbook,2ndbook}
can be used similarly for solving the multiplier determining equations. 
This reduces the problem of finding all conservation laws to a kind of 
adjoint of the problem of finding all symmetries. 
As an important consequence, 
all conservation laws admitted by a normal PDE system can be found 
in a straightforward computational way, without any need for special ansatzes or restrictions,
analogously to how all admitted symmetries can be found. 

For any given PDE system, 
there is a well-known natural action of its symmetries 
on all of its conservation laws
\cite{Olv,KarMah00,BluTemAnc}. 
This action allows conservation laws to be divided into 
symmetry equivalence classes, 
which can be used to define a generating subset (or a basis) \cite{KarMah02} 
for the set of all conservation laws of the PDE system. 

The present paper will explore a further connection 
between symmetries and conservation laws 
by focusing on conservation laws that are invariant 
(or, more generally, homogeneous) 
under the action of a given set of symmetries.
Some applications of symmetry-invariant conservation laws will also be discussed.
For simplicity, 
PDE systems consisting of a single equation for one dependent variable $u$
and two independent variables $t,x$ will be considered. 
All of the results have a direct generalization \cite{AncKar,review}
to normal PDE systems with more variables and more equations. 

In \secref{prelim}, 
the multiplier method for finding the conservation laws of a given PDE 
is reviewed, along with the action of symmetries on conservation laws. 
In \secref{formula}, 
a general formula \cite{AncBlu96,AncBlu97} that generates 
conservation laws from symmetries and adjoint-symmetries 
is connected to the infinitesimal action of a symmetry on a conservation law. 
One new result here will be to show that this formula is equivalent to 
a recent conservation law formula of Ibragimov \cite{Ibr07,Ibr10,Ibr11}. 

In \secref{results}, 
the notion of symmetry invariance of a conservation law will be defined and studied. 
This main result will yield a direct condition for invariance (and homogeneity) 
formulated in terms of multipliers.
In \secref{remarks}, 
some applications to finding symmetry-invariant conservation laws 
and finding symmetry-invariant solutions of PDEs 
will be outlined.

\section{Multipliers, conservation laws, and symmetries}
\label{prelim}

Consider an $N$th-order PDE 
\begin{equation}\label{pde}
G(t,x,u,\p u,\ldots,\p^N u)=0
\end{equation}
with dependent variable $u$, 
and independent variables $t,x$, 
where $\p^l u$ denotes all $l$th order partial derivatives of $u$ with respect to $t$ and $x$. 
The PDE \eqref{pde} is {\em normal} if 
it can be expressed in a solved form for some leading derivative of $u$ 
such that all other terms in the equation contain 
neither the leading derivative nor its differential consequences. 
All typical PDEs arising in physical applications are normal, \eg/ 
diffusion equations and dispersive wave equations, 
which have $u_t$ as a leading derivative; 
non-dispersive wave equations and hyperbolic equations, 
which have $u_{tt}$ or $u_{tx}$ as a leading derivative;
peakon equations and wave-breaking equations, 
which have $u_{txx}$ as a leading derivative. 

A {\em conservation law} \cite{Olv,2ndbook} of a given PDE \eqref{pde} is 
a total space-time divergence expression that vanishes on the solution space $\E$ of the PDE, 
\begin{equation}\label{conslaw}
(D_t T(t,x,u,\p u,\ldots,\p^r u) +D_x X(t,x,u,\p u,\ldots,\p^r u))|_\E=0. 
\end{equation}
The physical meaning of a conservation law is that 
the function $T$ is a conserved density 
while the function $X$ is a spatial flux. 
The pair 
\begin{equation}\label{current}
(T,X)=\Phi 
\end{equation}
is called a {\em conserved current}. 
Throughout, $D=(D_t,D_x)$ denotes a total derivative:
\begin{equation}
\begin{aligned}
D_t & =\p_t + u_t\p_u + u_{tx}\p_{u_x} + u_{tt}\p_{u_t} +\cdots , 
\\
D_x & =\p_x + u_x\p_u + u_{xx}\p_{u_x} + u_{tx}\p_{u_t} +\cdots . 
\end{aligned}
\end{equation}

Every conservation law \eqref{conslaw} can be integrated over 
any given spatial domain $\Omega\subseteq\Rnum$
\begin{equation}
\frac d{dt} \int_{\Omega} T dx = -X\Big|_{\p\Omega}
\end{equation}
showing that the rate of change of the quantity 
\begin{equation}\label{C}
\mathcal C[u]= \int_{\Omega} T dx
\end{equation}
is balanced by the net flux through the domain endpoints $\p\Omega$. 
Two conservation laws are physically equivalent if they give the same 
conserved quantity up to endpoint terms. 
This happens iff their conserved densities
differ by a total space derivative $D_x\Theta(t,x,u,\p u,\ldots,\p^r u)$
on the solution space $\E$,
and correspondingly, 
their fluxes differ by a total time derivative $-D_t\Theta(t,x,u,\p u,\ldots,\p^r u)$. 
A conservation law is called {\em locally trivial} if 
\begin{equation}\label{trivconslaw}
T|_\E = D_x\Theta,
\quad
X|_\E = - D_t\Theta
\end{equation}
so any two equivalent conservation laws differ by a locally trivial conservation law. 
For a given PDE \eqref{pde}, 
the set of all non-trivial conservation laws (up to equivalence)
forms a vector space. 

An {\em infinitesimal symmetry} \cite{Olv,1stbook,2ndbook} 
of a given PDE \eqref{pde} is a generator 
\begin{equation}\label{symm}
\X=\tau(t,x,u,\p u,\ldots,\p^r u)\p_{t} +\xi(t,x,u,\p u,\ldots,\p^r u)\p_{x} +\eta(t,x,u,\p u,\ldots,\p^r u)\p_{u}
\end{equation}
whose prolongation $\pr\X$ leaves invariant the PDE, 
\begin{equation}\label{inveq}
\pr\X(G)|_\E =0 . 
\end{equation}
When acting on the PDE solution space $\E$, 
any infinitesimal symmetry \eqref{symm} is equivalent to 
a generator with the {\em characteristic form} 
\begin{equation}\label{symmchar}
\hat\X=P\p_{u}, 
\quad 
P =\eta-\tau u_t-\xi u_x
\end{equation}
where the characteristic functions $\eta$, $\tau$, $\xi$ are determined by 
\begin{equation}\label{symmdeteq}
0 = \pr\hat\X(G)|_\E = G'(P)|_\E . 
\end{equation}
Here a prime denotes the Frechet derivative with respect to $u$:
\begin{equation}\label{frechet}
f'(g) = f_u g + f_{u_t}D_t g + f_{u_x}D_x g + f_{u_{tt}}D_t{}^2 g + f_{u_{tx}}D_tD_x g + f_{u_{xx}}D_x{}^2 g + \cdots
\end{equation}
for any differential functions $f(t,x,u,\p u,\p^2 u,\ldots)$
and $g(t,x,u,\p u,\p^2 u,\ldots)$. 
Note that the determining equations \eqref{inveq} and \eqref{symmdeteq}
are equivalent due to the identity $\pr\hat\X = \pr\X -\tau D_t -\xi D_x$. 

An infinitesimal symmetry of a given PDE \eqref{pde} is {\em trivial} 
if its action on the solution space $\E$ of the PDE is trivial, 
$\hat\X u =0$ for all solutions $u(t,x)$. 
This occurs iff $P|_\E=0$. 
The corresponding generator \eqref{symm} of a trivial symmetry 
is thus given by 
$\X|_\E= \tau\p_{t} + \xi\p_{x} +(\tau u_t +\xi u_x)\p_{u}$, 
which has the prolongation $\pr\X|_\E=\tau D_t + \xi D_x$. 
Conversely, any generator of this form on the solution space $\E$ 
determines a trivial symmetry. 

Since the infinitesimal symmetries of a PDE preserve its solution space, 
there consequently is a natural action by 
any symmetry on the set of all conservation laws of the PDE. 
The following result, or equivalent versions of it, are well known
\cite{Olv,KarMah00,BluTemAnc}. 

\begin{prop}\label{symmaction}
The action of an infinitesimal symmetry \eqref{symm} on a conserved current \eqref{current} 
is given by $\Phi_\X = \pr\X(\Phi) + \Phi D\cdot(\tau,\xi) -\Phi\cdot D(\tau,\xi)$. 
In explicit form, 
\begin{equation}
T_\X = \pr\X(T) + TD_x\xi - XD_x\tau,
\quad
X_\X = \pr\X(X) + XD_t\tau - TD_t\xi. 
\end{equation}
When the symmetry is expressed in the characteristic form \eqref{symmchar}, 
its action is simply given by $\Phi_{\hat\X} = \pr\hat\X\Phi$,
which has the explicit form 
\begin{equation}\label{symmactionTX}
T_{\hat\X} = \pr\hat\X(T)=T'(P), 
\quad
X_{\hat\X} = \pr\hat\X(X)=X'(P). 
\end{equation}
The conserved currents $\Phi_\X$ and $\Phi_{\hat\X}$ are equivalent, 
due to the relation $\pr\X-\pr\hat\X=\tau D_t + \xi D_x$. 
\end{prop}

An important observation is that the symmetry action on a given conserved current 
can produce a locally trivial conserved current \eqref{trivconslaw}. 

To determine when a conserved current is locally trivial, 
and to formulate determining equations for conserved currents, 
it is useful to have a characteristic (canonical) form for conservation laws.
The following results \cite{Olv,2ndbook,review} will be needed. 

\begin{lem}\label{hadamard}
If a differential function $f(t,x,u,\p u,\p^2 u,\ldots)$ vanishes 
on the solution space $\E$ of a given PDE \eqref{pde}, 
then $f= R_f(G)$ holds identically,
where $R_f =R_f^{(0)} + R_f^{(1)}\cdot D + R_f^{(2)}\cdot D^2 +\cdots $ 
is a linear differential operator, depending on $f$, 
with coefficients $R_f^{(i)}(t,x,u,\p u,\p^2 u,\ldots)$ that are non-singular on $\E$ whenever the PDE is normal. 
\end{lem}

\begin{lem}\label{totaldiv} 
A differential function $f(t,x,u,\p u,\p^2 u,\ldots)$ is a total space-time divergence $f=D_t A +D_x B$ for some functions
$A(t,x,u,\p u,\p^2 u,\ldots)$ and $B(t,x,u,\p u,\p^2 u,\ldots)$ 
iff $E_u(f)=0$ holds identically,
where $E_u = \p_u -D_t\p_{u_t}-D_x\p_{u_x} + D_t{}^2\p_{u_{tt}} + D_x{}^2\p_{u_{xx}} + D_tD_x\p_{u_{tx}} +\cdots$
is the Euler-Lagrange operator. 
\end{lem}

From \lemref{hadamard},
a conservation law $(D_t T +D_x X)|_\E=0$ for a normal PDE $G=0$
can be expressed as a divergence identity
\begin{equation}\label{conslawoffE}
D_t T +D_x X= R_\Phi(G)
\end{equation}
where $u(t,x)$ is an arbitrary (sufficiently differentiable) function. 
In this identity, integration by parts on the terms 
$R_\Phi(G)=
R_\Phi^{(0)}G + R_\Phi^{(1)t}D_tG + R_\Phi^{(1)x}D_xG + R_\Phi^{(2)tt}D_t{}^2G + R_\Phi^{(2)tx}D_tD_xG + R_\Phi^{(2)xx}D_x{}^2G + \cdots$ 
yields 
\begin{equation}\label{chareqn}
D_t\tilde T +D_x\tilde X= QG
\end{equation}
with 
\begin{equation}\label{equivTX}
\begin{aligned}
\tilde T  & = T - R_\Phi^{(1)t}G - R_\Phi^{(2)tt}D_t G +(D_tR_\Phi^{(2)tt})G - R_\Phi^{(2)tx}D_x G + \cdots,
\\\tilde X & = X - R_\Phi^{(1)x}G - R_\Phi^{(2)xx}D_x G +(D_xR_\Phi^{(2)xx})G +(D_tR_\Phi^{(2)tx})G + \cdots, 
\end{aligned}
\end{equation}
and
\begin{equation}\label{Q}
Q = R_\Phi^{(0)} - D_t R_\Phi^{(1)t} -D_x R_\Phi^{(1)x} +D_t{}^2R_\Phi^{(2)tt} +D_tD_xR_\Phi^{(2)tx} +D_x{}^2R_\Phi^{(2)xx} + \cdots . 
\end{equation}
On the PDE solution space $\E$, 
note that $\tilde T|_\E = T$ and $\tilde X|_\E = X$ reduce to 
the conserved density and the flux in the conservation law 
$(D_t T +D_x X)|_\E=0$, 
and hence $(D_t\tilde T +D_x\tilde X)|_\E= 0$ is an equivalent conservation law.
The identity \eqref{chareqn} is called the {\em characteristic equation}
for the conservation law, 
and the function \eqref{Q} is called the {\em multiplier}. 

A conserved density $T$ and a flux $X$ will determine a unique multiplier 
if the leading derivative of $u$ and all of its differential consequences
are first eliminated from $T$ and $X$ through the PDE $G=0$. 
However, the form of the multiplier expression \eqref{Q} shows that 
it can possibly depend on the leading derivative of $u$ 
as well as differential consequences of this derivative. 
In general a function $Q(t,x,u,\p u,\p^2 u,\ldots,\p^r u)$ will be a multiplier 
iff its product with the PDE $G=0$ has the form of 
a total space-time divergence. 

From the characteristic equation \eqref{chareqn}, 
it is straightforward to see that the multiplier \eqref{Q} 
for any locally trivial conservation law \eqref{trivconslaw} vanishes on the PDE solution space, $Q|_\E=0$.
Conversely, 
for any multiplier of the form $Q|_\E=0$, 
it can be shown \cite{review} that the characteristic equation \eqref{chareqn}
implies the conserved density and the flux are locally trivial \eqref{trivconslaw}.
This leads to the following basic result. 

\begin{thm}\label{correspondence}
For a normal PDE \eqref{pde}, 
there is a one-to-one correspondence between conservation laws (up to equivalence) and multipliers evaluated on the solution space of the system. 
\end{thm}

From this result, 
the {\em differential order of a conservation law} is defined to be 
the smallest differential order among all equivalent conserved currents.
A conservation law for a normal PDE \eqref{pde} is said to be of {\em low order}
if all derivative variables $\p^k u$ that appear in its multiplier 
are related to some leading derivative of $u$ 
by differentiation with respect to $t,x$. 
(Note that, therefore, the differential order $r$ of the multiplier 
must be strictly less than the differential order $N$ of the PDE.)
This definition is motivated by the observation \cite{AncKar,review} that 
physically important conservation laws, such as energy and momentum, 
are always of low order,
whereas higher order conservation laws are typically connected with integrability. 

For any normal PDE \eqref{pde}, 
all conservation law multipliers 
can be determined from \lemref{totaldiv} applied to 
the characteristic equation \eqref{chareqn}. 
This yields 
\begin{equation}\label{multdeteq}
0=E_u(QG)=G'{}^*(Q)+ Q'{}^*(G)
\end{equation}
which holds identically.
Here a star denotes the adjoint of the Frechet derivative with respect to $u$:
\begin{equation}\label{adjfrechet}
f'{}^*(g) = f_u g -D_t(f_{u_t} g) -D_x(f_{u_x} g) + D_t{}^2(f_{u_{tt}} g) + D_tD_x(f_{u_{tx}} g) + D_x{}^2(f_{u_{xx}} g) + \cdots
\end{equation}
for any differential functions $f(t,x,u,\p u,\p^2 u,\ldots)$
and $g(t,x,u,\p u,\p^2 u,\ldots)$. 
On the solution space $\E$ of the PDE \eqref{pde}, 
the determining equation \eqref{multdeteq} implies
\begin{equation}\label{adjsymmdeteq}
G'{}^*(Q)|_\E =0 . 
\end{equation}
From \lemref{hadamard}, it follows that $Q$ satisfies the identity 
\begin{equation}\label{adjsymmdeteqoffE}
G'{}^*(Q) = R_Q(G) . 
\end{equation}
Then the determining equation \eqref{multdeteq} becomes
\begin{equation}\label{helmholtzoffE}
0=R_Q(G)+ Q'{}^*(G)
\end{equation}
where $u(t,x)$ is an arbitrary (sufficiently differentiable) function. 
Since the PDE $G=0$ is assumed to have a solved form in terms of a leading derivative of $u$, 
the equation \eqref{helmholtzoffE} can be split with respect to 
this derivative of $u$ and its differential consequences, 
yielding a linear system of equations 
\begin{equation}\label{helmholtzeq}
(R_Q+ Q'{}^*)^{(i)} =0,
\quad
i=0,1,\ldots
\end{equation}
obtained from the coefficients in the linear differential operator $R_Q+ Q'{}^*$. 

Consequently, 
the determining equation \eqref{multdeteq} for multipliers is equivalent to 
the linear system of equations \eqref{helmholtzeq} and \eqref{adjsymmdeteq}. 
In this system, 
the first equation \eqref{adjsymmdeteq} is the adjoint of 
the symmetry determining equation \eqref{symmdeteq},
and its solutions $Q(t,x,u,\p u,\p^2 u,\ldots)$ 
are called {\em adjoint-symmetries} \cite{AncBlu97,AncBlu02a,AncBlu02b}. 
The remaining equations \eqref{helmholtzeq} comprise Helmholtz conditions \cite{Olv}
which are necessary and sufficient for $Q$ 
to have the form \eqref{Q} where $\Phi=(T,X)$ is a conserved current. 
Note that these conditions are equivalent to 
\begin{equation}\label{helmholtzadjeq}
(R_Q^*+ Q')^{(i)} =0,
\quad
i=0,1,\ldots
\end{equation}
which comes from expressing the equations \eqref{helmholtzeq} 
in the operator form $R_Q+ Q'{}^*=0$ and taking its adjoint. 

This formulation of a determining system for multipliers 
has a simple adjoint relationship to Noether's theorem, 
as will now be outlined. 
First, recall the condition for a PDE \eqref{pde} to be an Euler-Lagrange equation \cite{Olv,AncBlu96,1stbook,2ndbook}. 

\begin{prop} 
A PDE $G=0$ is an Euler-Lagrange equation $G=E_u(L)$ 
for some Lagrangian $L(t,x,u,\p u,\p^2 u,\ldots)$ iff $G'=G'{}^*$. 
\end{prop}

The following result is now straightforward to establish (see \Refs{AncBlu97,AncBlu02a,AncBlu02b}).

\begin{thm}\label{adjnoether}
For any normal PDE \eqref{pde}, 
conservation law multipliers are adjoint-symmetries \eqref{adjsymmdeteq}
that satisfy Helmholtz conditions \eqref{helmholtzeq}. 
In the case when the PDE is an Euler-Lagrange equation, 
adjoint-symmetries are the same as symmetries,
and the Helmholtz conditions are equivalent to symmetry invariance of the
Lagrangian modulo a total space-time divergence,
whereby multipliers for a Lagrangian PDE are the same as variational symmetries.
\end{thm}

It is important to note that the determining system for multipliers 
can be solved computationally by the same standard procedure \cite{Olv,1stbook,2ndbook}
used to solve the determining equation for symmetries. 
In particular, the multiplier determining system is more overdetermined 
than is the symmetry determining equation,
and hence the computation of multipliers is easier than the computation of symmetries. 

In applications of Theorem~\ref{adjnoether}, 
the use of a Lagrangian to obtain the conserved current from a variational symmetry 
is replaced by 
either a homotopy integral formula \cite{Olv,AncBlu02a,AncBlu02b}, 
or direct integration of the characteristic equation \cite{Wol02b,2ndbook},
both of which are applicable for any normal PDE. 
If a given PDE possesses a scaling symmetry 
then any conserved current having non-zero scaling weight 
can be obtained from an algebraic formula \cite{Anc} in terms of a multiplier 
(see also \Refs{DecNiv,PooHer}). 
Recently \cite{review}, 
this formula has been generalized to PDEs without a scaling symmetry,
by using scaling transformations that arise from dimensional analysis.

\section{A conservation law formula using symmetries and adjoint-symmetries}
\label{formula}

There is a general formula that relates symmetries and conserved currents
in an interesting way. 
Consider the Frechet derivative identity \cite{Olv,AncBlu96,AncBlu97,AncBlu02a,AncBlu02b,2ndbook}
\begin{gather}
hf'(g)- gf'{}^*(h) = {D\cdot\Psi_f(g,h)}
\label{frechetid}\\
\begin{aligned}
\Psi_f(g,h) = & 
h f_{\p u} g + h f_{\p^2 u}\cdot D g -(D\cdot(h f_{\p^2 u})) g 
+ h f_{\p^3 u}\cdot D^2 g 
\\&\qquad
-(D\cdot(hf_{\p^3 u}))D g +(D^2\cdot(hf_{\p^3 u})) g +\cdots
\end{aligned}
\label{divergenceid}
\end{gather}
where $f(t,x,u,\p u,\p^2 u,\ldots)$, $g(t,x,u,\p u,\p^2 u,\ldots)$ and $h(t,x,u,\p u,\p^2 u,\ldots)$ 
are arbitrary differential functions. 
The explicit form of the components of $\Psi_f=(\Psi_f^t,\Psi_f^x)$ 
is easily obtained from the Frechet derivative formula \eqref{frechet} and its adjoint \eqref{adjfrechet},
which gives 
\begin{equation}
\begin{aligned}
\Psi_f^t(g,h) = & 
g( h f_{u_t} -D_t(h f_{u_{tt}}) -D_x(h f_{u_{tx}}) +\cdots )
\\&\qquad
+ D_t g( h f_{u_{tt}} -D_t(h f_{u_{ttt}}) -D_x(h f_{u_{ttx}}) +\cdots )
\\&\qquad
+ D_t{}^2 g( h f_{u_{ttt}} -D_t(h f_{u_{tttt}}) -D_x(h f_{u_{tttx}}) +\cdots )
+\cdots
\end{aligned}
\end{equation}
modulo a trivial term $D_x\Theta$,
and
\begin{equation}
\begin{aligned}
\Psi_f^x(g,h) = & 
g( h f_{u_x} -D_x(h f_{u_{xx}}) +D_x{}^2(h f_{u_{xxx}}) +\cdots )
\\&\qquad
+ D_t g( h f_{u_{tx}} -D_x(h f_{u_{txx}}) +D_x{}^2(h f_{u_{txxx}}) +\cdots )
\\&\qquad
+ D_x g( h f_{u_{xx}} -D_x(h f_{u_{xxx}}) +D_x{}^2(h f_{u_{xxxx}}) +\cdots )
+\cdots
\end{aligned}
\end{equation}
modulo a trivial term $-D_t\Theta$. 

As first shown in \Refs{AncBlu96,AncBlu97}, 
the identity \eqref{frechetid} yields a conserved current by
putting $f=G$, $g=P$, $h=Q$, 
where $P$ is the characteristic of an infinitesimal symmetry \eqref{symmchar}
and $Q$ is an adjoint-symmetry, 
which satisfy $G'(P)|_\E=0$ and $G'{}^*(Q)|_\E=0$. 
Thus
\begin{equation}\label{PQconslawid}
D_t\Psi_G^t(P,Q)+D_x\Psi_G^x(P,Q)
= QG'(P) - PG'{}^*(Q) 
\end{equation}
vanishes on the solution space $\E$ of the given PDE $G=0$. 
The resulting conserved current $\Psi_G(P,Q)$ 
turns out to be directly related to the current 
$\Phi=(\tilde T,\tilde X)$ given by the characteristic equation 
$D_t\tilde T +D_x\tilde X=QG$
when $Q$ is a multiplier of a conservation law \eqref{chareqn}. 

\begin{prop}\label{symmactionrelation}
For a normal PDE \eqref{pde}, 
a conserved current is given by 
\begin{equation}\label{PQcurrent}
\Phi=(\Psi_G^t(P,Q),\Psi_G^x(P,Q))
\end{equation}
in terms of any given infinitesimal symmetry \eqref{symmchar}
and any given conservation law multiplier \eqref{chareqn}. 
The corresponding conservation law 
\begin{equation}\label{PQconslaw}
(D_t\Psi_G^t(P,Q)+D_x\Psi_G^x(P,Q))|_\E=0
\end{equation}
is equivalent to the conservation law
\begin{equation}\label{symmactionconslaw}
(D_t T'(P) +D_x X'(P))|_\E=0
\end{equation}
obtained from the action of the symmetry $\hat\X=P\p_u$ 
on the given conservation law whose multiplier is given by $Q$. 
In particular, the multipliers of these two conservation laws 
\eqref{PQconslaw} and \eqref{symmactionconslaw} 
are the same, 
$Q_{\Psi_G(P,Q)}=Q_{\hat\X}= R_P^*(Q) -R_Q^*(P)$. 
\end{prop}

This result has a straightforward proof by comparing the multipliers 
for the two conservation laws. 
(See \Ref{AncBlu96} for the variational case.)
It will be useful first to express the symmetry determining equation \eqref{symmdeteq} as an identity 
\begin{equation}\label{symmdeteqoffE}
\pr\hat\X(G) = G'(P) = R_P(G) 
\end{equation}
through \lemref{hadamard}. 
Now, consider the conservation law $(D_t\Psi_G^t(P,Q)+D_x\Psi_G^x(P,Q))|_\E=0$. 
Its multiplier can be obtained from the identity \eqref{PQconslaw} 
combined with the determining equations 
\eqref{symmdeteqoffE} and \eqref{adjsymmdeteqoffE} 
for symmetries and adjoint-symmetries.
This gives 
\begin{equation}
\begin{aligned}
D_t\Psi_G^t(P,Q)+D_x\Psi_G^x(P,Q)
& = Q R_P(G) - P R_Q(G)
\\
& = (R_P^*(Q)- R_Q^*(P))G + \text{ trivial current }
\end{aligned}
\end{equation}
after an integration by parts. 
Hence the multiplier is given by 
\begin{equation}\label{PQmultiplier}
Q_{\Psi_G(P,Q)} = R_P^*(Q)- R_Q^*(P) . 
\end{equation}
Next, consider the conservation law given by the characteristic equation
$D_t T + D_x X = QG$. 
From \propref{symmaction}, 
the action of the symmetry $\hat\X=P\p_u$ on this conservation law,
combined with the symmetry determining equation \eqref{symmdeteqoffE}, 
yields
\begin{equation}\label{PQchareqn}
\begin{aligned}
\pr\hat\X(D_t T +D_x X) 
& = D_t T'(P) +D_x X'(P) 
\\
& =\pr\hat\X(QG) =  Q'(P)G +QG'(P) =  Q'(P)G +QR_P(G)
\\
& =  (Q'(P) +R_P^*(Q))G + \text{ trivial current }
\end{aligned}
\end{equation}
after an integration by parts. 
Use of the Helmholtz equations \eqref{helmholtzadjeq} 
satisfied by multipliers gives the relation $Q'(P) + R_Q^*(P)=0$,
so thus equation \eqref{PQchareqn} becomes
\begin{equation}
\begin{aligned}
\pr\hat\X(D_t T +D_x X) 
& = D_t T'(P) +D_x X'(P) 
\\
& =  (R_P^*(Q)-R_Q^*(P) )G + \text{ trivial current }.
\end{aligned}
\end{equation}
Hence the multiplier for the conservation law 
$(D_t T'(P) +D_x X'(P))|_\E=0$ is given by 
\begin{equation}\label{symmactionQ}
Q_{\hat\X} = R_P^*(Q) -R_Q^*(P)
\end{equation}
which is the same as the multiplier $Q_{\Psi_G(P,Q)}$. 
This completes the proof. 

The explicit form of the conserved current \eqref{PQcurrent}, 
modulo locally trivial terms, is given by 
\begin{align}
\begin{aligned}
\Psi_G^t(P,Q) = & 
P( Q G_{u_t} -D_t(Q G_{u_{tt}}) -D_x(Q G_{u_{tx}}) +\cdots )
\\&\qquad
+ D_t P( Q G_{u_{tt}} -D_t(Q G_{u_{ttt}}) -D_x(Q G_{u_{ttx}}) +\cdots )
\\&\qquad
+ D_t{}^2 P( Q G_{u_{ttt}} -D_t(Q G_{u_{tttt}}) -D_x(Q G_{u_{tttx}}) +\cdots )
+\cdots , 
\end{aligned}
\label{PQ-Tformula}\\
\begin{aligned}
\Psi_G^x(P,Q) = & 
P( Q G_{u_x} -D_x(Q G_{u_{xx}}) +D_x{}^2(Q G_{u_{xxx}}) +\cdots )
\\&\qquad
+ D_t P( Q G_{u_{tx}} -D_x(Q G_{u_{txx}}) +D_x{}^2(Q G_{u_{txxx}}) +\cdots )
\\&\qquad
+ D_x P( Q G_{u_{xx}} -D_x(Q G_{u_{xxx}}) +D_x{}^2(Q G_{u_{xxxx}}) +\cdots )
+\cdots .
\end{aligned}
\label{PQ-Xformula}
\end{align}
These expressions first appeared in \Refs{AncBlu96,AncBlu97}
and are equivalent to a conservation law formula appearing later 
in work of Ibragimov \cite{Ibr07,Ibr10,Ibr11}, as will now be shown. 

The starting point is the standard observation \cite{Olv} 
that any PDE $G=0$ can be embedded into an Euler-Lagrange system 
by the introduction of an auxiliary dependent variable $v$, 
with the Lagrangian defined by $L=vG$. 
Then $E_v(L)=G=0$ and $E_u(L)=G'{}^*(v)=0$ are the Euler-Lagrange equations. 
Note that $G'{}^*(v)=0$ is simply a special case of 
the adjoint-symmetry equation \eqref{adjsymmdeteq} with $v=Q(t,x)$
(where the adjoint-symmetry $Q$ is restricted to have 
no dependence on $u$ and its derivatives). 
The Noether identity for this Lagrangian is given by 
$\pr\Y(L) = P^u G'{}^*(v) + P^v G + D\cdot\Psi_G(P^u,v)$ 
where $\Y= P^u\p_u + P^v\p_v$ is an arbitrary generator. 
Now consider any infinitesimal symmetry $\hat\X=P\p_u$ 
admitted by the PDE $G=0$.
From the symmetry identity \eqref{symmdeteqoffE}, 
it follows that $\pr\hat\X(L)= v R_P(G) = GR_P^*(v) + D\cdot\Phi_\triv$
after an integration by parts, 
where $\Phi_\triv|_\E=0$ is a locally trivial current. 
If the symmetry is extended to act on $v$ by $\hat\X^\ext=P\p_u - R_P^*(v)\p_v$,
then $\pr\hat\X^\ext (L)= D\cdot\Phi_\triv$ 
shows that this symmetry is variational.
Hence, Noether's theorem can be applied to obtain a conservation law
$D\cdot \Phi_{\X^\ext}=0$ 
on the solution space of the Euler-Lagrange equations $G=0$, $G'{}^*(v)=0$. 
The conserved current $\Phi_{\X^\ext}$ is obtained from the Noether identity
$\pr\hat\X^\ext (L)= D\cdot\Phi_\triv = P G'{}^*(v) -R_P^*(v)G + D\cdot\Psi_G(P,v)$,
which yields 
$D\cdot(\Phi_\triv-\Psi_G(P,v)) = P G'{}^*(v) -R_P^*(v)G$.
As a result, $\Phi_{\X^\ext}= \Psi_G(P,v)-\Phi_\triv$ is conserved 
on the solution space of the Euler-Lagrange equations $G=0$, $G'{}^*(v)=0$,
and therefore it defines a conserved current of the PDE $G=0$. 

All the steps in this derivation will go through if $v$ is allowed to be 
any adjoint-symmetry $Q(t,x,u,\p u,\ldots,\p^r u)$ admitted by the PDE $G=0$, 
in which case the conserved current defined by 
$\Phi_{\X^\ext}(P,Q)=\Phi_{\X^\ext}|_{v=Q}$ is equivalent to $\Psi_G(P,Q)$.
This conserved current $\Phi_{\X^\ext}(P,Q)$ is precisely the formula in Ibragimov's work. 

The importance of \propref{symmactionrelation} is that it explains
how Ibragimov's formula is simply a re-writing of 
the standard action of a symmetry on a given conservation law 
but where the symmetry action is expressed 
in terms of the multiplier underlying the conservation law
rather than more transparently in terms of the conserved current components of the conservation law. 
Moreover, 
the ``nonlinearly self-adjoint'' condition $G'{}^*(v)=0$ used by Ibragimov 
is a re-statement of the requirement that a given PDE admits an adjoint-symmetry, 
which is a necessary condition for a PDE to admit a conservation law. 
Consequently, 
every PDE that admits at least one conservation law is automatically ``nonlinearly self-adjoint''.

As one application, 
it is simple to explain a number of observations that have arisen from
the recent use of the formula \eqref{PQ-Tformula}--\eqref{PQ-Xformula} 
and its equivalent version in \Refs{Ibr07,Ibr10,Ibr11}.

\begin{prop}\label{PQformula}
(i) For a given normal PDE \eqref{pde}, 
let $Q$ be the multiplier for a conservation law in which the components of
the conserved current $(T,X)$ have no explicit dependence on $t,x$. 
Then, using any translation symmetry $\X=a\p_t+b\p_x$, 
with characteristic $P=-(au_t+bu_x)$ where $a,b$ are constants, 
the conserved current $\Psi_G(P,Q)$ is locally trivial. 
(ii) For a given normal PDE \eqref{pde} that possesses a scaling symmetry
$\X=at\p_t+bx\p_x+cu\p_u$, where $a,b,c$ are constants, 
let $Q$ be the multiplier for a conservation law in which the components of
the conserved current $(T,X)$ are scaling homogeneous. 
Then, using the characteristic $P=cu-(atu_t+bxu_x)$ of the scaling symmetry, 
the conserved current $\Psi_G(P,Q)$ is equivalent to a multiple $w$ of 
the conserved current $\Phi$ determined by $Q$. 
This multiple $w$ is the scaling weight of the conserved integral 
$\int_{\Omega} T dx$ \cite{Anc}. 
\end{prop}

The proof uses the equivalent formula provided by \propref{symmactionrelation}. 
Let $(T,X)$ be a conserved current determined by a multiplier $Q$. 
For any translation symmetry $\X=a\p_t+b\p_x$, 
it is easy to see that 
$\Psi_G(P,Q) = (-T'(au_t+bu_x),-X'(au_t+bu_x))$ 
holds modulo a locally trivial current.
The Frechet derivative formula \eqref{frechet} gives the relation 
$f'(au_t+bu_x)= aD_t f+bD_x f -af_t -bf_x$ 
where $f$ is any differential function. 
Since both $T,X$ are assumed to have no dependence on $t,x$, 
it follows that $T'(au_t+bu_x)|_\E= (aD_t T+bD_x T)|_\E = D_x(bT-aX)$,
and similarly $X'(au_t+bu_x)|_\E= (aD_t X+bD_x X)|_\E = -D_t(bT-aX)$. 
This shows that $\Psi_G(P,Q)$ is a locally trivial current.
Next, for a scaling symmetry $\X=at\p_t+bx\p_x+cu\p_u$, 
it is straightforward to see that 
$\Psi_G(P,Q) = (T'(cu-(atu_t+bxu_x)),X'(cu-(atu_t+bxu_x)))$ 
holds modulo a locally trivial current.
If both $T,X$ are scaling homogeneous then it follows that 
$T'(cu-(atu_t+bxu_x))= pT -atD_t T-bxD_x T$
and $X'(cu-(atu_t+bxu_x))= qX -atD_t X-bxD_x X$, 
where $p,q$ are the respective scaling weights of $T,X$,
which satisfy the scaling relation $p-b=q-a$. 
This yields $(pT -atD_t T-bxD_x T)|_\E = (p-b)T +D_x((at+bx)X)$
and similarly $(qX -atD_t X-bxD_x X)|_\E = (p-b)X -D_t((at+bx)X)$. 
Hence $\Psi_G(P,Q)$ is equivalent to $(wT,wX)$, 
with $w=p-b$ being the scaling weight of $\int_{\Omega} T dx$. 
This completes the proof.

\section{Main results}
\label{results}

From \proprefs{symmaction}{symmactionrelation},
together with the identity $\pr\hat\X=\pr\X -\tau D_t-\xi D_x$
for any symmetry generator \eqref{symm}, 
the following main result is immediate. 

\begin{thm}\label{symmactionTXQ}
For a given normal PDE \eqref{pde},
let $\Phi=(T,X)$ be a conserved current and $Q$ be its multiplier. 
Then, for any infinitesimal symmetry \eqref{symm} admitted by the PDE, 
$\Phi_{\hat\X} = \pr\hat\X(\Phi)$ is also a conserved current
and $Q_{\hat\X} = R_P^*(Q) -R_Q^*(P)$ is its multiplier. 
In explicit form, 
\begin{equation}
\begin{aligned}
& T_{\hat\X} = T'(P) 
= \pr\X(T) + TD_x\xi  - XD_x\tau +D_x\Theta -(D_t T+D_x X)\tau , 
\\
& X_{\hat\X} = X'(P) 
= \pr\X(X) + XD_t\tau - TD_t\xi -D_t\Theta -(D_t T+D_x X)\xi , 
\end{aligned}
\end{equation}
on the PDE solution space $\E$, 
whereby the conserved current $\Phi_{\hat\X} = (T'(P),X'(P))$ 
is equivalent to the conserved current 
$\Phi_{\X}=(\pr\X(T) + TD_x\xi  - XD_x\tau,\pr\X(X) + XD_t\tau - TD_t\xi)$. 
These currents are locally trivial iff 
$Q_{\hat\X} = R_P^*(Q) -R_Q^*(P)=0$ vanishes identically. 
\end{thm}

The action of an infinitesimal symmetry on a multiplier, 
as well as the condition for this symmetry action to yield a locally trivial conserved current,
are new results. 

In \Ref{KarMah00}, 
the notion of associating a conserved current to a symmetry is introduced 
by considering the condition that a conserved current is invariant
under the action of an infinitesimal symmetry,
\begin{equation}
\Phi_{\X}=(\pr\X(T) + TD_x\xi  - XD_x\tau,\pr\X(X) + XD_t\tau - TD_t\xi) =0 . 
\end{equation}

A more appropriate notion of invariance is that a conserved current is 
equivalent to a locally trivial current under the action of an infinitesimal symmetry,
\begin{equation}\label{invconslaw}
\Phi_{\X}|_\E=(\pr\X(T) + TD_x\xi  - XD_x\tau,\pr\X(X) + XD_t\tau - TD_t\xi)|_\E 
=(D_x\Theta,-D_t\Theta) .
\end{equation}
This corresponds to invariance of the conservation law $(D_t T+D_x X)|_\E=0$,
which can be formulated in a simple way in terms of its multiplier $Q$, 
giving $Q_{\hat\X} = R_P^*(Q) -R_Q^*(P)=0$ 
as shown by combining \thmrefs{symmactionTXQ}{correspondence}.

Symmetry invariance of a conservation law has a natural extension 
by allowing a conserved current to be equivalent to a multiple of itself 
under the action of an infinitesimal symmetry, 
\begin{equation}\label{homoconslaw}
(\Phi_{\X} -\lambda\Phi)|_\E 
=(\pr\X(T) + TD_x\xi  - XD_x\tau-\lambda T,\pr\X(X) + XD_t\tau - TD_t\xi-\lambda X)|_\E + (D_x\Theta,-D_t\Theta) 
\end{equation}
where $\lambda$ is a non-zero constant. 
This leads to the following results, 
from \thmrefs{symmactionTXQ}{correspondence}.

\begin{thm}
A conservation law \eqref{conslaw} is homogeneous \eqref{homoconslaw}
under the action of an infinitesimal symmetry \eqref{symm} 
iff its multiplier \eqref{Q} satisfies the condition 
\begin{equation}\label{invcond}
R_P^*(Q) -R_Q^*(P)=\lambda Q
\end{equation}
for some constant $\lambda$. 
The conservation law is symmetry invariant \eqref{invconslaw} iff $\lambda=0$. 
If a PDE is an Euler-Lagrange equation $G=E_u(L)$ for some Lagrangian $L$, 
then every conservation law is invariant under the variational symmetry 
corresponding to its multiplier. 
\end{thm}

\begin{cor}
(i) 
Under the action of an infinitesimal symmetry \eqref{symm},
a conserved quantity \eqref{C} on a spatial domain $\Omega\subseteq\Rnum$ 
is unchanged modulo an arbitrary endpoint term,
$\hat\X C[u] = \Theta|_{\p\Omega}$, 
iff the corresponding conservation law \eqref{conslaw} is symmetry invariant. 
(ii) 
A conserved quantity \eqref{C} is mapped into itself 
(modulo an arbitrary endpoint term) under a symmetry 
iff the corresponding conservation law \eqref{conslaw} is symmetry homogeneous. 
In particular, 
under the action of a symmetry transformation $\exp(\epsilon\hat\X)$
with parameter $\epsilon$, 
a symmetry-homogeneous conserved quantity $C[u]$ is mapped into 
$\exp(\epsilon\lambda)C(u)$ whenever all endpoint terms vanish. 
\end{cor}

Examples of these results are presented in \Ref{AncKar} 
for several nonlinear PDEs arising in a variety of physical applications.

\section{Concluding remarks}
\label{remarks}

Conservation laws that are symmetry invariant or symmetry homogeneous 
have at least three important applications.
Firstly, 
each symmetry-homogeneous conservation law (up to equivalence) 
represents a one-dimensional invariant subspace 
in the set of all non-trivial conservation laws (modulo locally trivial conservation laws). 
This is a useful feature when a generating subset (or basis) is being sought.
Secondly, 
any symmetry-invariant conservation law will reduce to a first integral 
for the ODE obtained by symmetry reduction of the given PDE
when symmetry-invariant solutions $u(t,x)$ are sought. 
This provides a direct reduction of order of the ODE. 
Thirdly, 
the determining equations for multipliers can be augmented by 
the symmetry-homogeneity condition,
which allows symmetry-homogeneous conservation laws to be obtained 
in a direct way by solving the augmented determining system.

\section*{Acknowledgements}

S.C. Anco is supported by an NSERC research grant.

\end{document}